# Traffic experiment reveals the nature of car-following

Rui Jiang[1,2,*], Mao-Bin Hu[2], H.M.Zhang[3,4], Zi-You Gao[1], Bin-Jia[1], Qing-Song Wu[2],

Bing Wang[5], Ming Yang[5]

[1] MOE Key Laboratory for Urban Transportation Complex Systems Theory and Technology, Beijing Jiaotong University, Beijing 100044, China

[2] State Key Laboratory of Fire Science and School of Engineering Science, University of Science and Technology of China, Hefei 230026, China

[3] Department of Civil and Environmental Engineering, University of California Davis, CA 95616, United States

[4] Department of Traffic Engineering, School of Transportation Engineering, Tongji University, Shanghai 200092, China

[5] Department of Automaton, Shanghai Jiaotong University, Shanghai 200240, China

*Email: rjiang@ustc.edu.cn

**Abstract**: As a typical self-driven many-particle system far from equilibrium, traffic flow exhibits diverse fascinating non-equilibrium phenomena, most of which are closely related to traffic flow stability and specifically the growth/dissipation pattern of disturbances. However, the traffic theories have been controversial due to a lack of precise traffic data. We have studied traffic flow from a new perspective by carrying out large-scale car-following experiment on an open road section, which overcomes the intrinsic deficiency of empirical observations. The experiment has shown clearly the nature of car-following, which runs against the traditional traffic flow theory. Simulations show that by removing the fundamental notion in the traditional car-following models and allowing the traffic state to span a two-dimensional region in velocity-spacing plane, the growth pattern of disturbances has changed qualitatively and becomes qualitatively or even quantitatively in consistent with that observed in the experiment.

**Introduction**

Vehicular traffic congestion is a serious problem all over the world, which causes huge economic loss and environmental pollution. In the last few decades, traffic flow studies have attracted wide interests of scientists from various disciplines [1]-[6]. This is because, on the one hand, the studies are practically important to mitigate traffic congestion, and on the other hand, traffic flow is a self-driven many-particle system far from equilibrium. The competitive, nonlinear interactions among vehicles give rise to the formation of diverse fascinating phenomena such as boundary induced phase transitions, spontaneous formation of jams, metastability, hysteresis, phase-separation, and gridlock formation [7]-[11]. Therefore, the study of traffic flow offers an opportunity to understand various fundamental aspects of nonequilibrium systems which are of current interest in statistical physics, such as pedestrian traffic and panics [12][13], granular flow [14], bacterial colonies [15], swarming and herding of birds and fishes [16], traveling waves in an emperor penguin huddle [17], ant trails [18], movement of molecular motors along filaments [19], dynamic arrest in glasses [20], oscillations of stock market dynamics [21], and the "bull-whip" effect of supply chain [22].

Most of the nonequilibrium phenomena of uninterrupted traffic flow on highways or expressways are closely related to traffic flow stability and specifically the growth/dissipation pattern of disturbances. However, unfortunately, there have been many controversies in traffic flow research [23]-[27]. Some traffic engineers believe that all traffic jams are induced by bottlenecks [23]. When traffic jams occur for no apparent reason, which is known as a phantom jam, it is only because nobody has looked far enough to find the reason. On the other hand, two decades ago, some physicists found that traffic jams can arise completely spontaneously and no bottleneck is necessary [28]-[30]. Under the right conditions small and local fluctuations, which happen all the time on the roads, could trigger a non-local traffic congestion that persists for hours. Recently, experimental studies on traffic flow on a circular road demonstrated that traffic jams can occur on a road without bottlenecks [31]-[33].

In the traditional traffic flow theory, a fundamental notion is that there is a unique relationship between the flow rate and the traffic density (or equivalently between traffic speed and vehicle spacing) under steady state conditions [34]-[37]. A traffic flow might be stable, metastable, or unstable. Disturbances in the metastable and unstable traffic flows could grow and develop into oscillating traffic via a subcritical Hopf bifurcation, like limit cycles generated in nonlinear autonomous systems. The traffic flow system is thus characterized as a self-excited (autocatalytic) oscillator [35]. In contrast, in the three-phase traffic theory proposed by Kerner [38]-[43], it is supposed that the steady state of congested traffic occupies a two-dimensional region in the flow density plane. Kerner pointed out that there are three traffic phases: free flow, synchronized flow, and wide moving jams. Usually phase transitions from free flow to synchronized flow occur firstly. At the upstream of the synchronized flow, there exists a pinch region, in which pinch effect induces small jams. The small jams propagate towards upstream, grow, merge, and develop into wide moving jams.

The controversy between the three-phase traffic flow theory and the traditional one is still on-going, mainly due to a lack of precise traffic data [24]-[27]. Although there are vast amounts of empirical data collected by loop detectors, video cameras and floating cars, we are unfortunately still not able to know exactly how the traffic flow evolves even in very simple scenarios, such as in a platoon of passenger cars following each other without overtaking and led by a car moving with constant velocity. This is because, on the one hand, as claimed by Daganzo *et al.*, "no empirical studies to date describe the complete evolution of a disturbance" and thus we cannot "explain the genesis of the disturbances or the rate at which they grow" [23]. On the other hand, the empirical observations are quite site-specific and contain many confounding factors (e.g., geometry, bottleneck strength, traffic flow composition), which prevents us from forming a comprehensive understanding of traffic flow evolution.

Motivated by this problem, we conducted controlled car-following experiments to examine the traffic flow theories. Different from the previous experiments carried out on a circular road [31]-[33], our experiments concern a platoon of 25 passenger cars on a 3.2-km-long open road section. Since the location and velocity of each individual car have been recorded by high precision GPS devices, the complete evolution of the disturbances and their growth rate can be obtained. Moreover, in the experiments, (i) the influence of road geometries has been excluded; (ii) The leading car has been asked to move with different constant speed so that various traffic flow situations can be studied, which goes beyond in-situ observations of traffic flow on roads. Therefore, general conclusions about car-following behavior are expected.

We also would like to mention that in microscopic traffic flow modeling, one usually firstly establishes and studies the car-following models, which is the basis of the microscopic modeling and plays an extremely important role in traffic flow studies. Then one introduces lane changing behaviors and other measurements into the car-following models to deal with multilane traffic flow, mixed traffic flow, road geometries etc. Our experimental results therefore provide a base dataset to examine the car-following models and such data are almost impossible to obtain from empirical observations.

**Experimental setup**

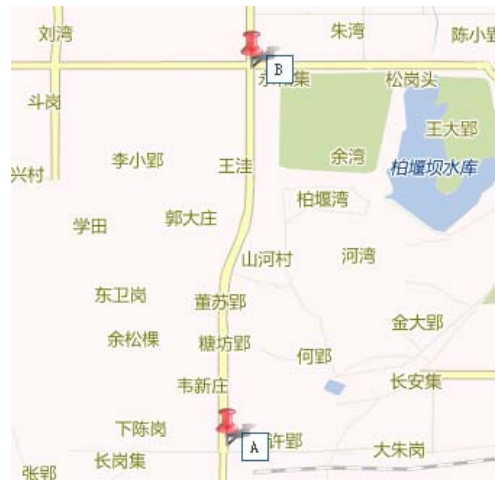

**Figure 1.   Map of the 3.2 km long experiment road section, which is between points A and B**

The experiment was carried out on January 19, 2013 on a 3.2 km stretch of the Chuangxin Avenue in a suburban area in Hefei City, China. See Figure 1 for the map of the section. There are no traffic lights on the road section. Since the road is located in a suburban area and has at least three lanes in each direction, there is no interference from other vehicles that are not part of the experiment.

High-precision GPS devices were installed on all of the cars to record their locations and velocities every 0.1 second. The measurement errors of the GPS devices were within ± 1 m for location and within ± 1 km/h for velocity. Once the experiment starts, the driver of the leading car is required to control the velocity of the car at certain pre-determined constant values. Other drivers in the experiment are required to drive their cars as they normally do, but overtaking is not allowed. In two of the runs, the driver of the leading car is required to firstly drive the car at 15 km/h as accurate as he can, then to drive the car without pressing the accelerator pedal (the velocity is about 7 km/h in this case), and then to drive the car at 20 km/h. In other runs, the driver of the leading car is required to accelerate the car to a pre-determined velocity and then maintain the car at the constant velocity as accurate as he can. The leading car is equipped with a cruise control system, which could be switched on when its velocity shown by the speedometer reaches 45 km/h. As a result, the fluctuation of the leading car is very small when its velocity is set at a value greater than or equal to 45 km/h. We note that for safety reason, the actual velocity of a car is lower than that shown by the speedometer, in particular when the velocity is high. See Figure 2 for the examples of the velocity of the leading car. When reaching the end of the road section, the car platoon decelerates, makes U-turn, and stops. When all the cars have stopped, a new run of the experiment begins.

Due to the length limit of the road section, we only analyze the experiment data in which the speed of the leading car is below or equal to 50 km/h. When the leading car moves with speed larger than 50 km/h, the experimental data are too limited after removing the data belonging to the start process, the deceleration and U-turn process. We have carried out two sets of experiments, in which the sequence of the cars has been changed, see Table 1.

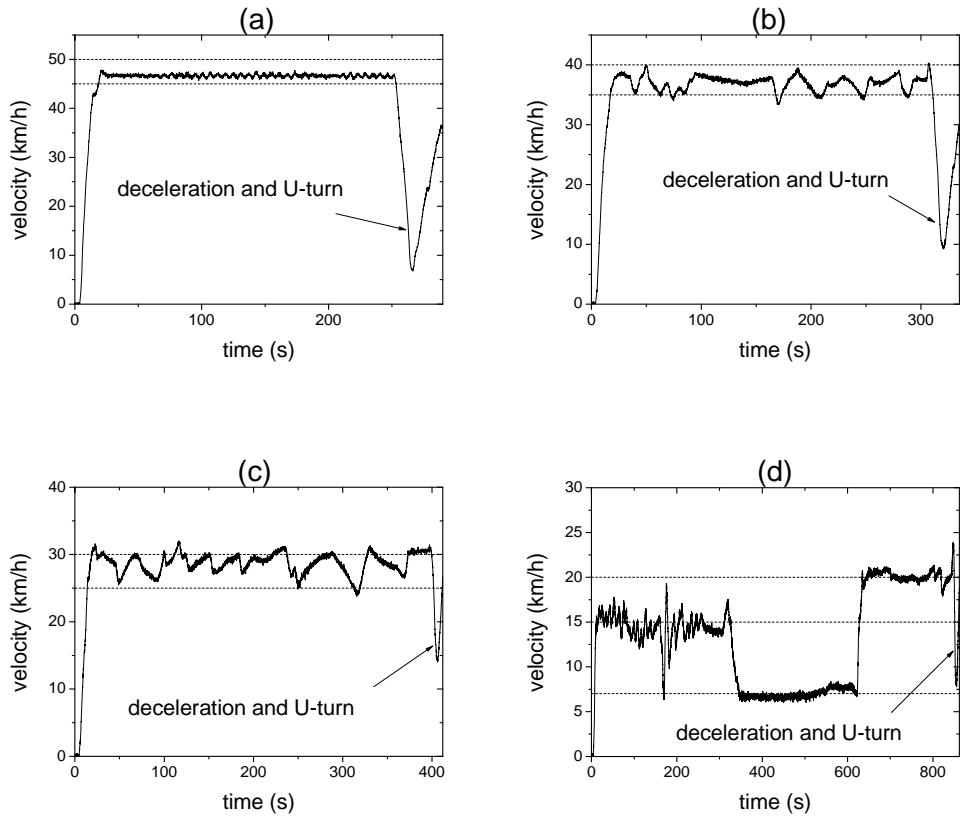

**Figure 2. The velocity of the leading car.** The driver of the leading car is asked to drive at (a) 50 km/h. (b) 40 km/h. (c) 30 km/h. (d) firstly at 15 km/h, then to drive the car without pressing the accelerator pedal (the velocity is about 7 km/h in this case), and then to drive the car at 20 km/h. Note that for safety reason, the actual velocity of a car is lower than that shown by the speedometer, in particular when the velocity is high. The dashed lines are guide for the eyes.

Table 1 Sequence of the cars in the two sets of the experiment.

| Sequence of Car in the platoon | Car No. in the 1st set of experiment | Car No. in the 2nd set of experiment |
|---|---|---|
| 1 | 22 | 22 |
| 2 | 1 | 11 |
| 3 | 2 | 12 |
| 4 | 3 | 13 |
| 5 | 4 | 15 |
| 6 | 5 | 16 |
| 7 | 6 | 17 |
| 8 | 7 | 18 |
| 9 | 8 | 19 |
| 10 | 9 | 20 |
| 11 | 10 | 21 |
| 12 | 11 | 23 |
| 13 | 12 | 25 |
| 14 | 13 | 1 |
| 15 | 14 | 2 |
| 16 | 15 | 3 |
| 17 | 16 | 4 |
| 18 | 17 | 5 |
| 19 | 18 | 6 |
| 20 | 19 | 7 |
| 21 | 20 | 8 |
| 22 | 21 | 9 |
| 23 | 23 | 10 |
| 24 | 24 | 14 |
| 25 | 25 | 24 |

**Results**

**Car-following behavior**

Figures 3(a) and 3(b) show two examples of the evolution of the spacing of car No.1 (which is the second car in the platoon, see Table 1) as well as the velocities of the car and its preceding car in the first set of experiment. In Figure 3(a), the car velocity is nearly constant and around 25 km/h, and the velocity difference between the car and its preceding one is always small. Before t = 174 s, the spacing fluctuates slightly around 15 m. The fundamental notion in traditional traffic flow theory could be correct, provided the observations before t =174s is the norm. However, after t =174 s, the spacing begins to fluctuate significantly and has increased to more than 30 m, even though the traveling speed of that car remains nearly constant. Figure 3(b) shows that the spacing also changes significantly between 15 m and 40 m when the cars move with velocity around 41 km/h. Figures 3(c)-(f) show some more examples of other cars, and similar results are observed.

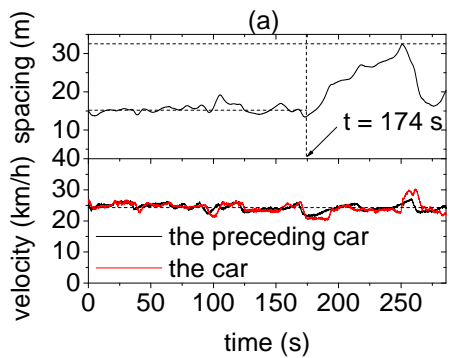
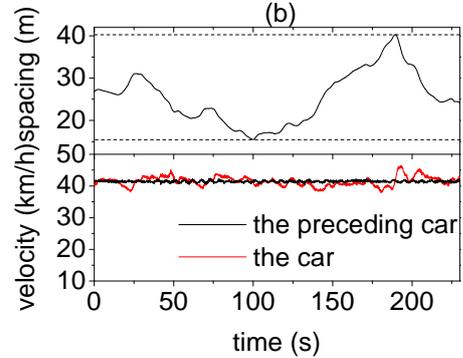
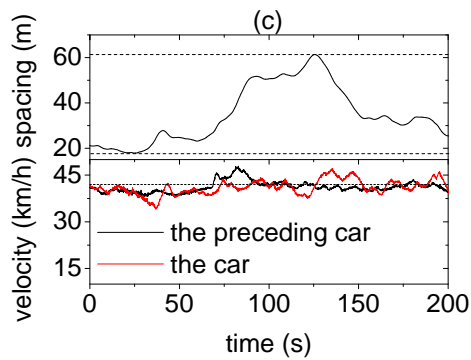
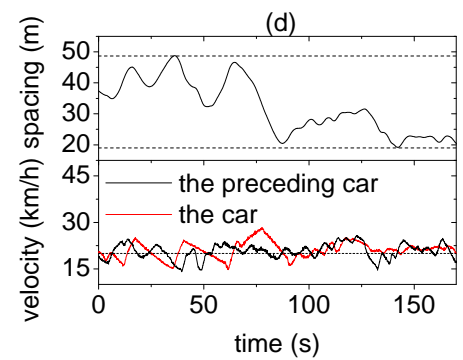

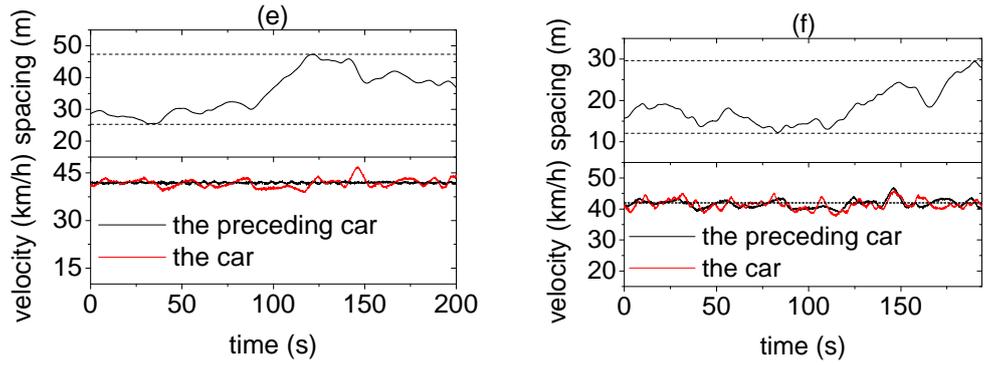

**Figure 3. Evolution of the spacing of a car as well as the velocities of the car and its preceding car.** (a) and (b) Car No.1 in the 1st set of experiment. (c) Car No.2 in the 1st set of experiment. (d) Car No.23 in the 1st set of experiment. (e) Car No.11 in the 2nd set of experiment. (f) Car No.12 in the 2nd set of experiment. See Table 1 for the sequence of the cars in the platoon in the two sets of experiment.

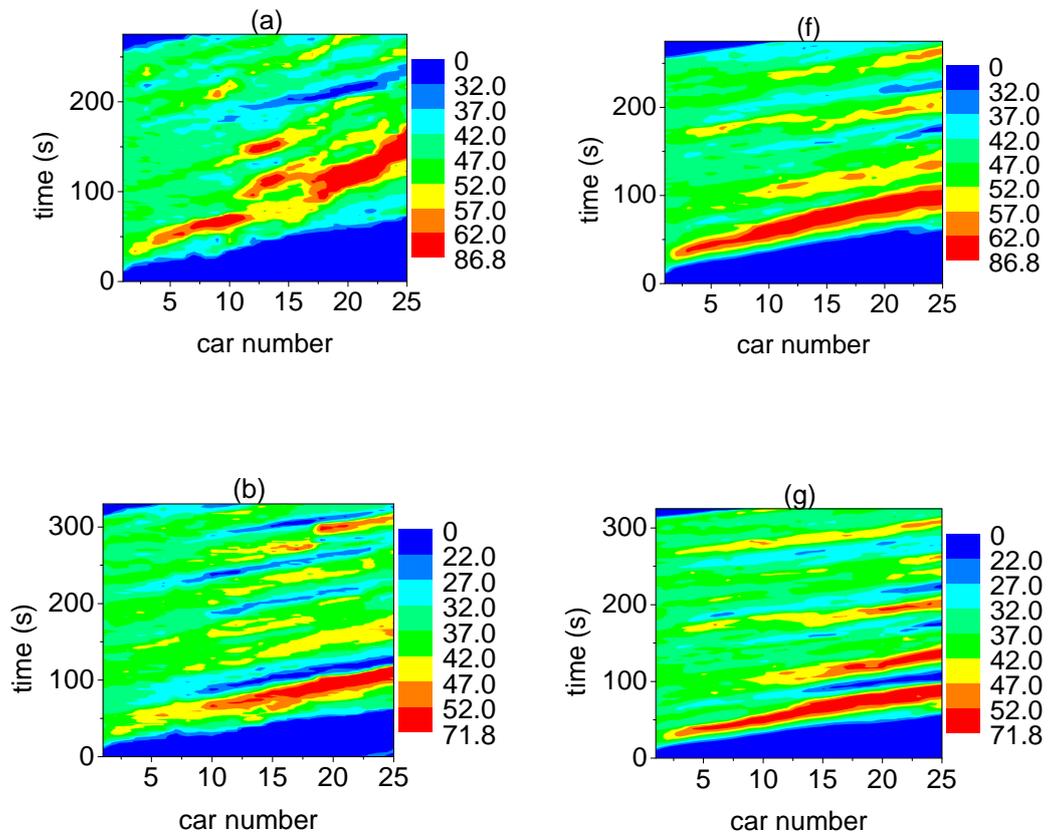

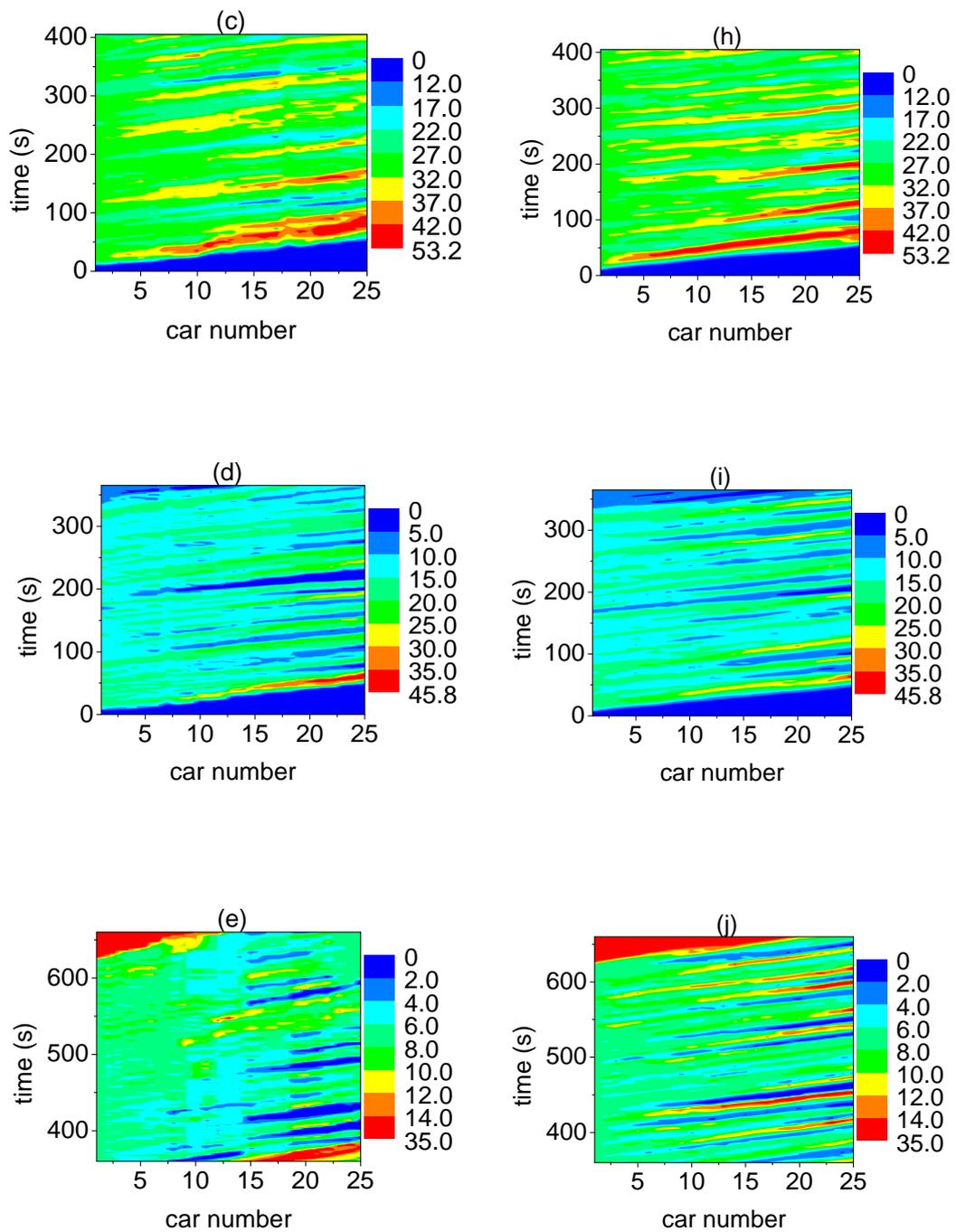

**Figure 4. Typical spatiotemporal patterns of the platoon traffic: car speed shown with different colors (unit: km/h) as function of time and car number.** The left column shows the experimental results and the right column shows the simulation results of the 2D intelligent driver model. (a)-(e), The velocity of the

leading car is shown in Figure 2. (f)-(j), In the simulation, the velocity of the leading car is set the same as in the experiment.

This feature clearly contradicts the fundamental assumption in traditional traffic flow theory. Here are two possible explanations of this feature. (i) In certain range of spacing, drivers are not so sensitive to the changes in spacing when the velocity differences between cars are small. Only when the spacing is large (small) enough, will they accelerate (decelerate) to decrease (increase) the spacing. (ii) At a given velocity, drivers do not have a fixed preferred spacing. Instead they change their preferred spacing either intentionally or unintentionally from time to time in the driving process.

**Spatiotemporal features**

Figures 4(a)-(e) show some typical spatiotemporal patterns of the traffic flow. One can observe the stripes structure, which exhibits the propagation, growth, dissipation and merge of disturbances. In Figures 4(a)-(d), in which the leading car is not very slow, the velocity fluctuation amplitude of the cars in the rear part of the platoon has exceeded 20 km/h. In Figure 4(d), in which the leading car moves with velocity around 15 km/h, some cars in the rear part of the platoon have occasionally completely stopped. In Figure 4(e), in which the leading car moves very slowly with velocity around 7 km/h, the cars in the rear part of the platoon will have to completely stop from time to time.

To better understand the evolution of the disturbances, Figure 5 presents the standard deviation $\sigma_v$ of the time series of the velocity of each car over all runs of the experiment, in which the leading car is asked to move with the same velocity. Note that the data belonging to the start process, the deceleration and U-turn process have been removed. One can see that when the velocity $v_l$ of the leading car is low, $\sigma_v$ increases almost linearly with car number, see Figure 5(a). With the increase of $v_l$, $\sigma_v$ becomes increasing in a concave way, i.e., the curve of $\sigma_v$ bends downward and the increment rate decreases, see Figures 5(b)-(d).

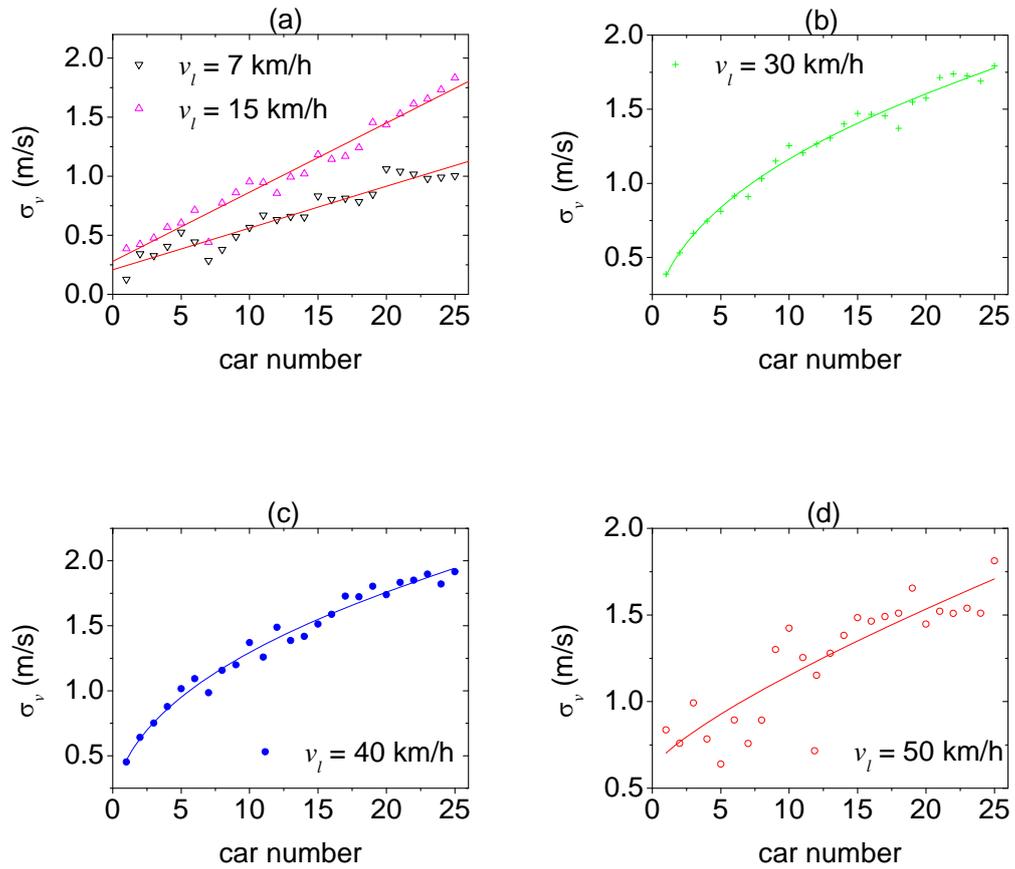

**Figure 5. The standard deviation $\sigma_v$ of the time series of the velocity of each car over all runs of the experiment, in which the leading car is asked to move with the same velocity $v_l$.** The solid lines suggest possible curve fits.

**Discussion**

**Simulations with traditional models**

To study if the observed traffic features of the 25-car platoon can be reproduced with traditional car-following theories, we choose four traditional car-following models to perform traffic simulation. The details of the models and the parameters used are as follows.

- The equation of the optimal velocity (OV) model [30]:

$$\frac{dv_i}{dt} = \kappa[V(\Delta x_i) - v_i] + \xi$$

Here $\Delta x_i = x_{i-1} - x_i$ is the spacing of car $i$, which follows vehicle $i-1$. $\kappa$ is sensitivity parameter. $\xi$ is noise to represent the stochastic factors in traffic flow, which is simply assumed to be uniformly distributed random number between $-\xi_1$ and $\xi_1$.

The OV function used is $V(\Delta x) = 11.6(\tanh(0.086(\Delta x - 25)) + 0.913)$ [44]. The unit of length and time are meter and second, respectively. The parameters are $\kappa = 1$ s$^{-1}$, $\xi_1 = 0.2$ m/s$^2$. The simulation time step $\Delta t = 0.1$ s

- The equation of the full velocity difference (FVD) model [45]:

$$\frac{dv_i}{dt} = \kappa[V(\Delta x_i) - v_i] + \lambda(v_{i-1} - v_i) + \xi$$

Here $\lambda$ is another sensitivity parameter. The OV function and the noise $\xi$ are the same as in the OV model. The parameters are $\kappa = 0.32$ s$^{-1}$, $\lambda = 0.4$ s$^{-1}$, $\xi_1 = 0.2$ m/s$^2$. The simulation time step $\Delta t = 0.1$ s

- The equation of the intelligent driver (ID) model [46]:

$$\frac{dv_i}{dt} = a\left[1 - \left(\frac{v_i}{v_{max}}\right)^4 - \left(\frac{s_0 + v_i T + \frac{v_i(v_i - v_{i-1})}{2\sqrt{ab}}}{\Delta x_i - l}\right)^2\right] + \xi$$

Here $a$ is maximum acceleration, $b$ is desired deceleration, $T$ is desired time headway, $s_0$ is the desired gap (bumper-to-bumper distance) between two neighboring cars in jam, $v_{max}$ is maximum velocity, $l$ is car length, $\xi$ is noise as before. The parameters are $v_{max}$ = 80 km/h, $T$ = 1.6 s, $a$ = 0.73 m/s², $b$ = 1.67 m/s², $s_0$ = 2 m, $l$ =5 m, $\xi_1$ = 0.2 m/s². The simulation time step $\Delta t = 0.1$ s

- The equation of the inertial model [47]:

$$\frac{dv_i}{dt} = A\left(1 - \frac{v_i T + D}{\Delta x_i}\right) - \frac{Z^2(v_i - v_{i-1})}{2(\Delta x_i - D)} - kZ(v_i - v_{per}) + \xi$$

Here $A$ is a sensitivity parameter, $D$ is the minimal distance between consecutive cars, $v_{per}$ is the permitted velocity, $k$ is a constant, $T$ is the safety time gap constant, $\xi$ is noise as before. The function $Z$ is defined as $Z(x) = (|x| + x)/2$. The parameters are $A$ = 5 m/s², $D$ = 5 m, $v_{per}$ = 80 km/h, $k$ = 2 s$^{-1}$, $T$ =2 s, $\xi_1$ = 0.2 m/s². The simulation time step $\Delta t = 0.1$ s

Figure 6 shows the curves of the standard deviation of the velocities of the cars. In the simulation, the leading car in an initially stopped platoon begins to accelerate with a constant acceleration until it reaches velocity $v_l$, then it maintains the velocity $v_l$. One can see that quantitatively the simulation results significantly deviate from the experimental ones. More importantly, the simulated curves increase in a convex way, which is qualitatively different from the experimental ones.

We would like to mention that since the velocity of a car is bounded by zero speed and the maximum speed, the standard deviation $\sigma_v$ cannot grow infinitely. Thus, the curve of $\sigma_v$ cannot always grow in a convex way.

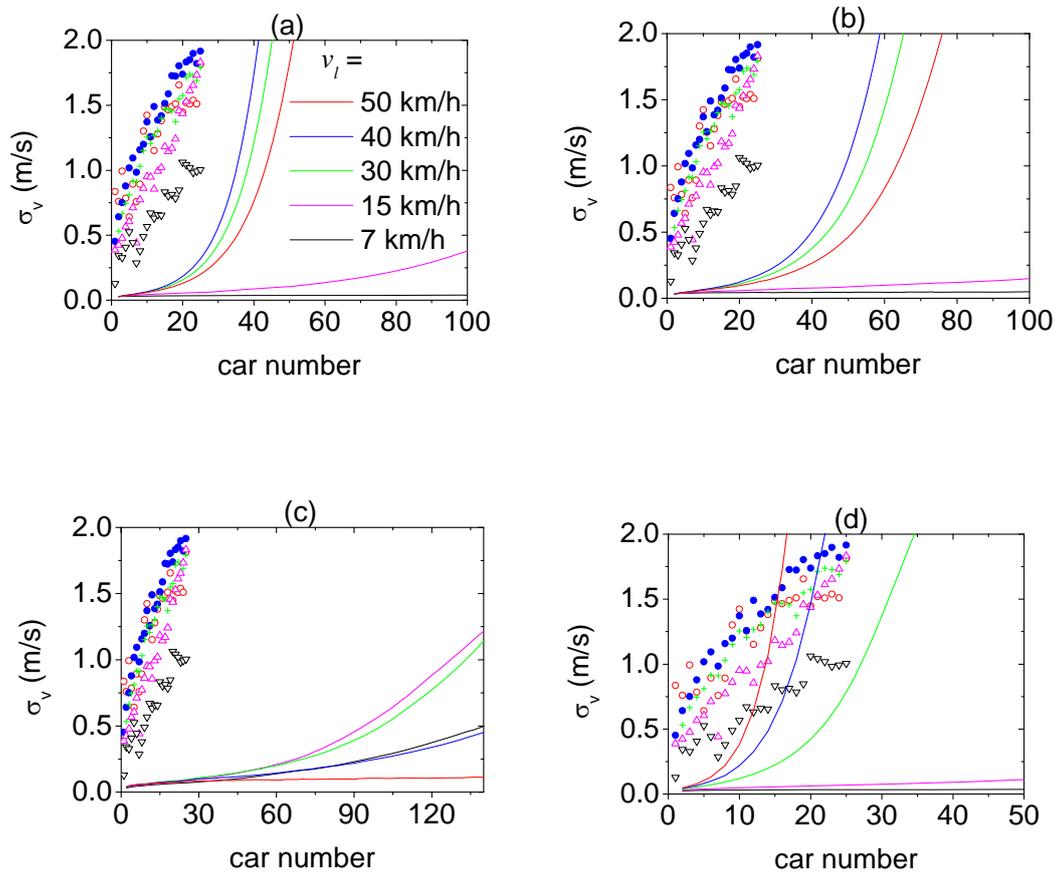

**Figure 6. Simulation results (solid lines) of the standard deviation $\sigma_v$ of the time series of the velocity of each car.** (a) the OV model, (b) the FVD model, (c) the ID model, (d) the inertial model. The simulation results are independent of the constant acceleration of the leading car. The scatter points are from the experiment.

**Simulations with new models**

We develop new models by removing the fundamental notion that there is a unique relationship between traffic speed and vehicle spacing in the steady state in the traditional models. As a result, the traffic state of a car could span a two-dimensional (2D) region in the velocity spacing plane. Thus, the new models are named as 2D models. The details of the 2D models and the parameters used are as follows.

- The 2D OV and FVD models:

We assume that the relationship between speed and spacing is determined by $V(\Delta x) = \max(11.6(\tanh(0.086(m\Delta x - 25)) + 0.913), 0)$ with a parameter $m$, which changes with time. For simplicity, we assume that $m$ is a uniformly distributed random number between $m_1$ and $m_2$, and $m$ changes in the range with rate $p$ (i.e., in each time step $\Delta t$ in the simulation, $m \rightarrow m'$ with probability $p\Delta t$ and remains unchanged with probability $1 - p\Delta t$). In the simulations, the parameters are $p = 0.15$ s$^{-1}$, $m_1 = 0.8$, $m_2 = 1.2$. Other parameters are the same as before.

- The 2D ID model:

We suppose that the drivers do not always maintain a constant value of desired time headway $T$. For simplicity, we assume that $T$ is a uniformly distributed random number between $T_1$ and $T_2$, and $T$ changes in the range with rate $p$. In the simulation, the parameters are $T_1 = 0.5$ s, $T_2 = 1.9$ s, $p = 0.15$ s$^{-1}$. Other parameters are the same as before.

- The 2D inertial model:

Similar as in the 2D ID model, we assume that $T$ is a uniformly distributed random number between $T_1$ and $T_2$, and $T$ changes in the range with rate $p$. The parameters are $p = 0.15$ s$^{-1}$, $T_1 = 1.6$ s, and $T_2 = 2.4$ s. Other parameters are the same as before.

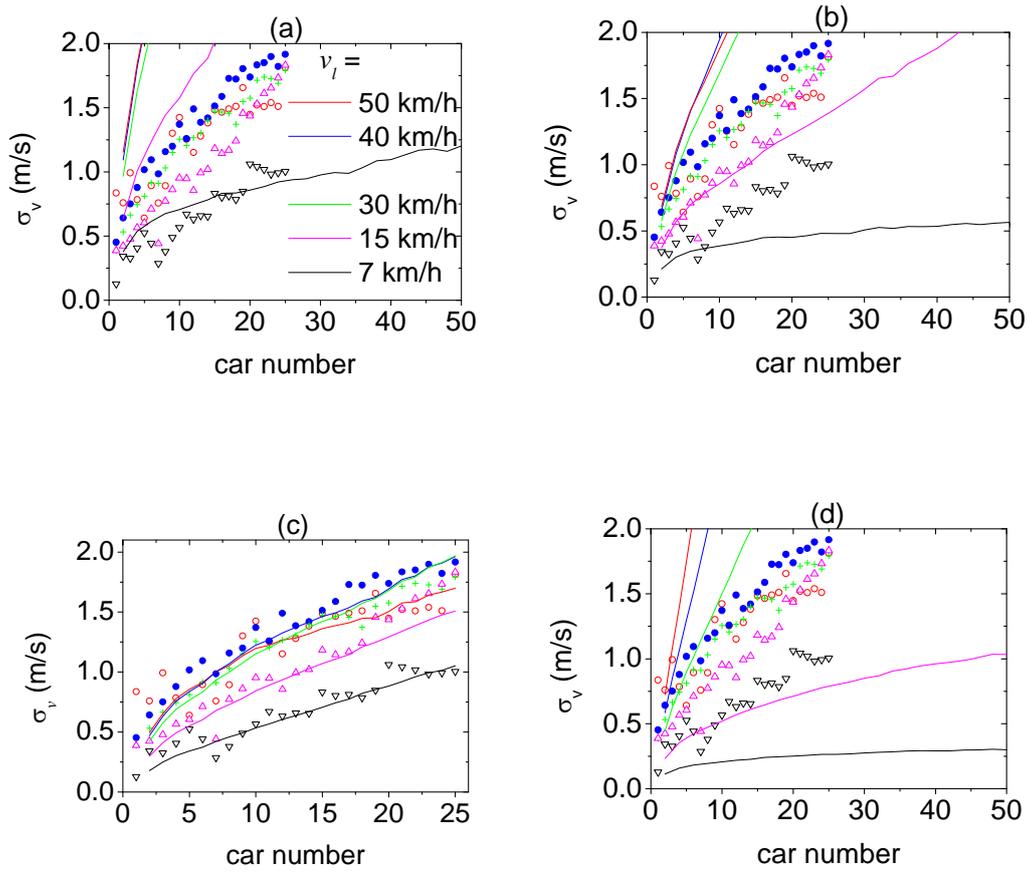

**Figure 7. Simulation results (solid lines) of the standard deviation $\sigma_v$ of the time series of the velocity of each car**. (a) the 2D OV model, (b) the 2D FVD model, (c) the 2D ID model, (d) the 2D inertial model. The scatter points are from the experiment.

The curves of the standard deviation of the velocities of the cars in the 2D models are shown in Figure 7. One can see that although the simulation results are still quantitatively different from the experimental ones in Figures 7(a), 7(b), and 7(d), the growth of the disturbances changes from a convex way into a concave way, which is qualitatively consistent with the experimental ones. This implies that the feature of traffic flow stability has qualitatively changed after removing the fundamental notion. As for the 2D ID model, the growth of the disturbances is not only qualitatively but also quantitatively in good agreement with the experimental ones, as shown in Figure 7(c). Figures 4(f)-(j) show the spatiotemporal patterns of the velocities simulated from the 2D ID model, in which formation of the stripes is very similar to those in the experiment.

**Conclusions**

To summarize, we have carried out large scale car-following experiments on an open section of a road. The experimental evidence clearly runs against the traditional traffic flow theory. Simulations with traditional car-following models reproduce qualitatively different growth pattern of disturbances from the experiment. We have developed new car-following models, in which the fundamental notion in traditional traffic flow theory is removed and the traffic state of a car could span a two-dimensional region in the velocity-spacing plane. The new models are able to qualitatively or even quantitatively reproduce the growth pattern of disturbances, which implies that the feature of traffic flow stability has changed qualitatively.

We would like to note that due to the limit of road section length, the duration of the stationary state is short when the traffic speed is high (only about 2 minutes when the leading car moves with the velocity of 55 km/h). Thus, the features of high-speed traffic flow have not been revealed. Therefore, in future work, larger-scale experiments on longer road sections and with larger platoon size need to be carried out to examine the issue.

car-following model. Phys Rev Lett 84: 382-385